\documentclass[prl,reprint,amsmath,amssymb,twocolumn,superscriptaddress,showpacs]{revtex4}

\usepackage{graphicx}
\usepackage{dcolumn}
\usepackage{bm}
\usepackage{times}
\usepackage{units}

\usepackage[pdftex]{color}

\newcommand\create[2]{\hat{#1}^\dagger_{#2}}
\newcommand\annihi[2]{\hat{#1}^{\phantom{\dagger}}_{#2}}

\begin{document}


\title{
Unconventional Superconductivity from Local Spin Fluctuations in the Kondo Lattice
}

\date{\today}
\author{Oliver Bodensiek}
\affiliation{Institut f\"ur Theoretische Physik, Universit\"at G\"ottingen, 37077
G\"ottingen, Germany}
\author{Rok \v{Z}itko}
\affiliation{J.\ Stefan Institute, Jamova 39, SI-1000 Ljubljana, Slovenia}
\author{Matthias Vojta}
\affiliation{Institut f\"ur Theoretische Physik, Technische Universit\"at Dresden, 01062 Dresden, Germany
}
\author{Mark Jarrell}
\affiliation{
Louisiana State University, Baton Rouge, Louisiana 70803, USA
}
\author{Thomas Pruschke}
\affiliation{Institut f\"ur Theoretische Physik, Universit\"at G\"ottingen, 37077
G\"ottingen, Germany}

\pacs{71.27.+a,74.20.-z,74.20.Mn}

\begin{abstract}
The explanation of heavy-fermion superconductivity is a long-standing challenge to
theory. It is commonly thought to be connected to non-local fluctuations of either spin
or charge degrees of freedom and therefore of unconventional type. 
Here we present results for the Kondo-lattice model, a
paradigmatic model to describe heavy-fermion compounds, obtained from dynamical mean-field theory which
captures local correlation effects only. Unexpectedly, we find robust $s$-wave superconductivity in the 
heavy-fermion state.
We argue that this novel type of pairing is tightly connected to the formation of
heavy quasiparticle bands and the presence of strong {\em local} spin fluctuations.
\end{abstract}

\maketitle


Heavy-fermion (HF) 4f and 5f intermetallic compounds constitute a paradigm for strong
electronic correlations. Their low-temperature behavior is affected by f-shell local
moments subject to antiferromagnetic (AF) exchange coupling to the conduction electrons,
resulting in Fermi-liquid (FL) phases with strongly renormalized Landau parameters, most
notably huge effective masses
\cite{stewart:84,GreweSteglich:1991,stewart:2001,hewson:book}.
HF materials often display symmetry-breaking phases which
occur either within the heavy FL \cite{stewart:84} or compete with it \cite{loehneysen:2007}.
While magnetic order in systems containing unscreened moments appears natural,
HF superconductivity is conceptually non-trivial and indeed came as an unexpected
discovery more than three decades ago \cite{steglich:1979}.
By now, a wide variety of f-electron superconductors are known
\cite{Thalmeier_Zwicknagl:2005,pfleiderer_rmp}, many of them confirmed to be
unconventional \cite{stewart:84,GreweSteglich:1991,Thalmeier:2005}.

Superconducting (SC) transitions in HF compounds are often assumed to be driven by
non-local fluctuations of the f-shell spin degrees of freedom. This idea finds support in
the close connection between HF superconductivity and magnetic quantum phase transitions
(QPTs) where such fluctuations are strong
\cite{stewart:2001,loehneysen:2007,nair:2010,stockert:2011}.
Alternatively, pairing mediated by fluctuations in the charge channel (i.e., valence
fluctuations) has also been discussed \cite{miyake:2007}.
Given that the basic theoretical models for HF materials, the periodic Anderson model (PAM) and
Kondo-lattice model (KLM), constitute complicated interacting many-body problems which
cannot be solved exactly,  theoretical descriptions of HF superconductivity often employ either
simple static mean-field theories or effective models of fermions coupled to spin
or charge fluctuations.

A rather successful approach to study the microscopic properties of
correlated-electron lattice models beyond static mean-field or effective descriptions 
is the dynamical mean-field theory
(DMFT) together with its cluster extensions \cite{georges:96,maier:05}. 
The lattice problem is mapped onto a self-consistent
quantum impurity model at the expense of losing information on
non-local correlation effects beyond the spatial size of the impurity
cluster.
Therefore, it is commonly assumed that a proper description of HF superconductivity
within DMFT-based approaches requires either large enough clusters or the inclusion of a
bath in the two-particle channel (which explicitly models a bosonic ``glue'' for
superconductivity). In particular, within the conventional (single-site) DMFT
only $s$-wave superconductivity occurs \cite{pruschke:95} and a relation to
magnetic fluctuations appears highly unlikely.

In this Letter we report on the unexpected observation of a stable SC solution to the
DMFT equations for the KLM \emph{without any external glue}. Although the pairing
symmetry is $s$-wave, the SC state is highly unconventional: Pairing is driven by {\em
local} spin fluctuations; it comes with a strong frequency dependence of the gap function
and requires the formation of HF bands as a prerequisite.

We note that a hint of the possible occurrence of local superconductivity was found in an
earlier DMFT study to the PAM \cite{Tahvildar-Zadeh:1998} which, however, did not analyze
the SC phase but only normal-state instabilities. Static mean-field descriptions of the
KLM or the PAM can also yield solutions with local pairing \cite{Howczak:2012,masuda:12,suppl}, but
it is difficult to assess their validity, as fluctuations beyond mean field may destroy
pairing.


\paragraph{Model.}
Within the KLM the localized f-states are
described by a (pseudo-)spin degree of freedom which couples to the
conduction ($c$) electrons
via an exchange interaction. We use the simplest version of the model, i.e., a
nearest-neighbor tight-binding conduction band with spin degeneracy only and a $S=1/2$
spin located at each lattice site. The Hamiltonian reads
\begin{equation}\label{eq:KLM}
\mathcal{H}=
-t\sum_{\langle i,j\rangle,\sigma}\create{c}{i\sigma} \annihi{c}{j\sigma}+
\frac{J}{2}\sum_{i,\alpha\beta}\hat{\mathbf{S}}_i\cdot   \,\create{c}{i\alpha}
 \boldsymbol{\tau}_{\alpha\beta}\annihi{c}{i\beta}.
\end{equation}
Here $\hat{c}_{i\sigma}^{(\dagger)}$ denote annihilation (creation) operators of
conduction electrons with spin $\sigma$ at site $i$, and $\langle.,.\rangle$ denotes
nearest-neighbors. $\hat{\mathbf{S}}_i$  is the operator for the
localized spin,
$\boldsymbol{\tau}$ the vector of Pauli matrices, and the interaction between the
conduction electrons and the localized spin is modeled as an isotropic exchange
coupling with $J>0$.
For simplicity, we will consider nearest-neighbor hopping $t$ on the infinite-dimensional
Bethe lattice, leading to a semi-circular density of states with bandwidth $W$. We have checked that using different
lattice types (for example hypercubic or square lattice tight-binding) does not change the results qualitatively.


\paragraph{Methods.}
Within standard DMFT, the conduction-electron self-energy is approximated as local in space,
$\Sigma(\mathbf{k},\omega) \rightarrow \Sigma(\omega)$. Then, the KLM maps onto an effective
single-impurity Kondo model (SIKM) \cite{hewson:book} which needs to be solved within a
self-consistency loop.
In this work we treat the effective SIKM using Wilson's numerical renormalization group (NRG) \cite{bullareview}.
It allows one to access arbitrarily small energy scales, to calculate spectra directly
on the real-frequency axis,
and to work at both $T=0$ and $T>0$. We work with the discretization
parameter $\Lambda=2.0$, keep $N_{st}=1000,\ldots,2000$ states, and perform $z$-averaging with $N_z=2$ \cite{bullareview}.

To allow for solutions with SC order, we generalize the DMFT equations and the impurity
solver to a Nambu formulation with $2\times 2$ matrix propagators
\cite{bullareview,Bauer:2009}.
This constrains our
calculations to spin-singlet even-frequency $s$-wave superconductivity. 
The DMFT treatment of superconductivity is non-perturbative and thus goes beyond the
standard Eliashberg theory \cite{Freericks:1994}: it does not rely on any assumption
about a separation of energy scales for the fermions and the bosonic glue responsible for
the formation of superconductivity.
At present, we restrict the calculations to SC order only, suppressing possible magnetic
order. Our result below show that strong pairing occurs in a regime without
magnetic order, justifying this neglect.


\begin{figure}[tb]
\includegraphics[width=.48\textwidth]{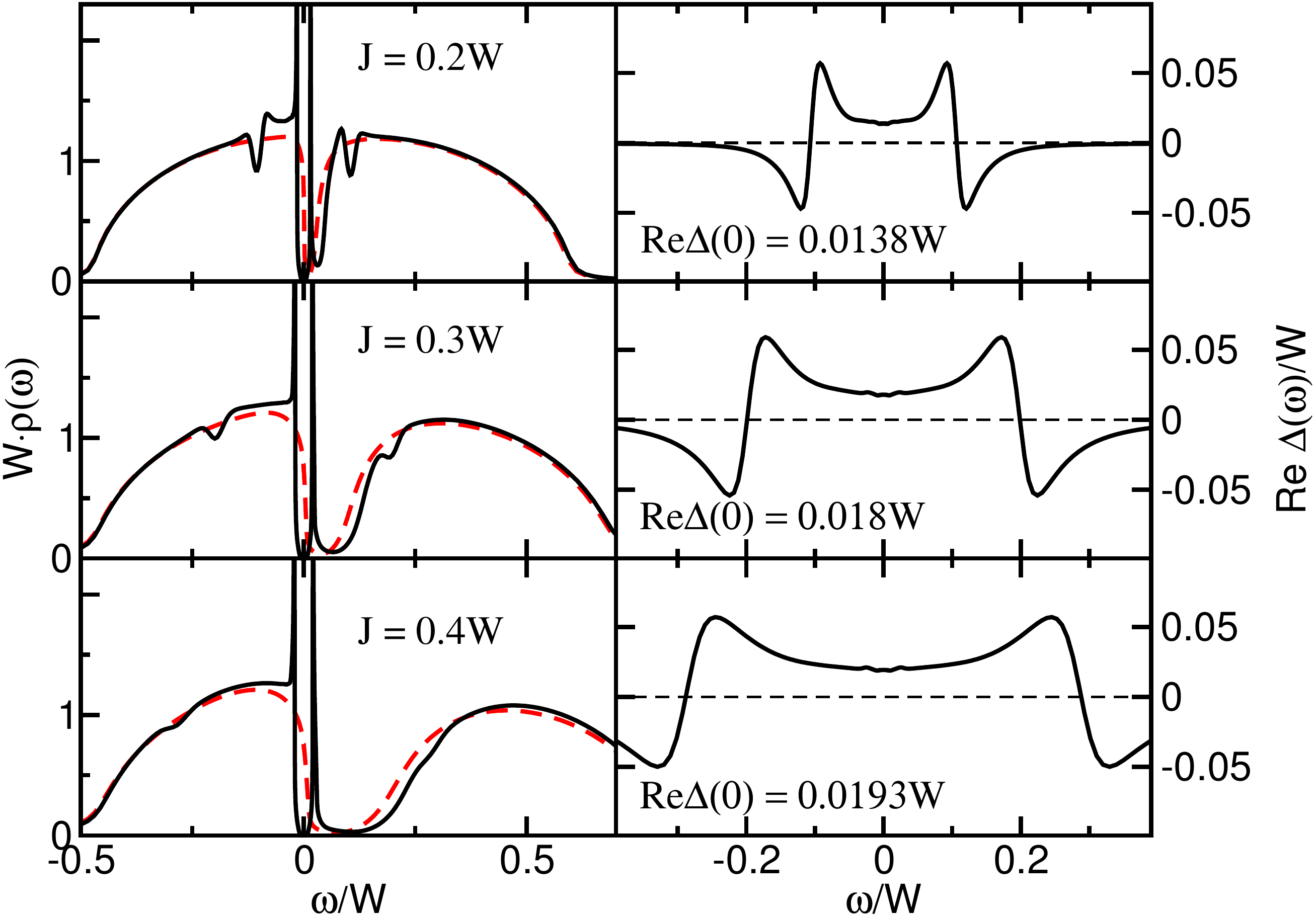}
\caption{(color online)
Left panel: N DOS (dashed/red lines) and SC DOS (full lines) for $n=0.9$ and various $J$. Right
panel: Real part of the corresponding gap functions.
}
\label{fig:KLdev}
\end{figure}


\paragraph{Results: Superconductivity at $T=0$.}
Our numerical solution of the DMFT equations yielded, for a range of
model parameters, stable SC solutions whose properties we discuss in the following.
The conduction-band density of states (DOS) of both the paramagnetic normal (N) and SC solutions of the
KLM for fixed conduction band filling $n=0.9$ and different $J$ are
shown in Fig.~\ref{fig:KLdev}, left panels.
The only feature of the N DOS is a
hybridization pseudo-gap above the Fermi energy, signaling the formation of heavy
quasiparticles, which can be rationalized within the picture of hybridized $c$ and $f$ bands.
The N solution is unstable  against superconductivity,
where the DOS exhibits two additional features:
(i) A true gap $\Delta_{\text{sc}}$ with well-developed van-Hove singularities is present around the Fermi
energy; as a function of $J$, it first increases up to $J/W=0.5$, and then slowly decreases.
(ii) In addition to the SC coherence peaks, there are side resonances
at positions which roughly scale with $J$. These structures are sharp for smaller $J$, but
become increasingly washed out for larger $J$.
These features are likely related to local spin fluctuations stabilizing the pairing, as
discussed in the supplemental material. 

As the appearance of a gap alone is not sufficient to identify the solution as a SC, one needs to 
look at the anomalous parts of the Nambu Green's function respectively the anomalous part of the
self-energy. From it, a SC gap function can be defined, as in standard Eliashberg analysis, via
\begin{align}
\label{eq:gapfunc}
\Delta(\omega) = \frac{\Sigma_1(\omega)+i \Sigma_2(\omega)}{1-\Sigma_0(\omega)/\omega},
\end{align}
where $\Sigma_\alpha(\omega)$ denote  the components of the electronic
self-energy expanded into Pauli matrices, $\Sigma = \Sigma_\alpha\tau_\alpha$
($\alpha=0,1,2,3$), in Nambu space.
The resulting real parts Re$\Delta(\omega)$ are shown in the right
panels of
Fig.~\ref{fig:KLdev}. As expected for even-frequency pairing, Re$\Delta(\omega)$ is
symmetric. It shows a strong frequency dependence, with sharp features shifting to larger
energies and broadening with increasing $J$. These structures are linked to the side
resonances in the DOS: the zeroes of Re$\Delta$ coincide with
the resonances in the DOS. 
The $\omega=0$ limit, Re$\Delta(0)$, provides an estimate of the gap seen in the DOS; it
exhibits the same non-monotonic behavior with $J$ as noted above for the gap in the DOS.

\begin{figure}[!b]
\includegraphics[width=0.5\textwidth]{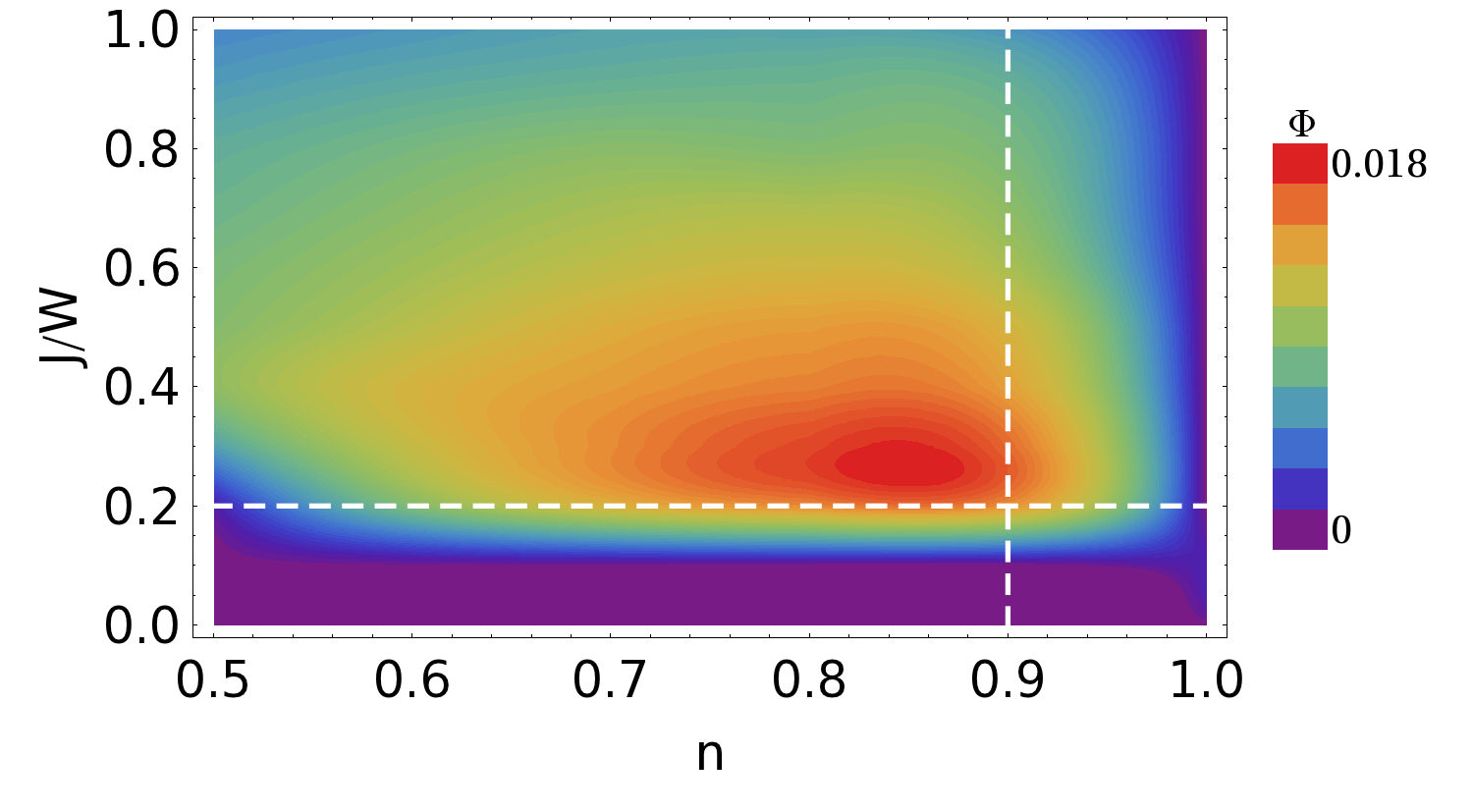}
\caption{(color online)
Anomalous expectation value $\Phi$ as a function of $J/W$ between quarter and half filling. The 
white dashed lines indicate the cuts along a fixed $J$ resp. $n$ shown in Fig.~\ref{fig:klorder} (a), (b).\\}
\label{fig:KLPD}
\vspace{-0.5cm}
\end{figure}
To characterize the evolution of superconductivity across the phase diagram, we plot
in Fig.~\ref{fig:KLPD} the anomalous expectation value $\Phi=\langle
\hat{c}_{i\uparrow}\hat{c}_{i\downarrow}\rangle$, as a function of $J$ and $n$. 
Superconductivity is found to be stable over wide regions of the phase diagram for $J/W > 0.1$.
For a fixed $J/W=0.2$, a finite $\Phi$ is found between $0.45 < n < 1$. For larger $J/W$ the 
SC region extends to even lower fillings. A maximum of $\Phi$ appears 
around $J/W = 0.3$ and $n=0.9$; the
side-resonances are also the most pronounced for these parameters.

\begin{figure}[!t]
    \includegraphics[width=0.48\textwidth]{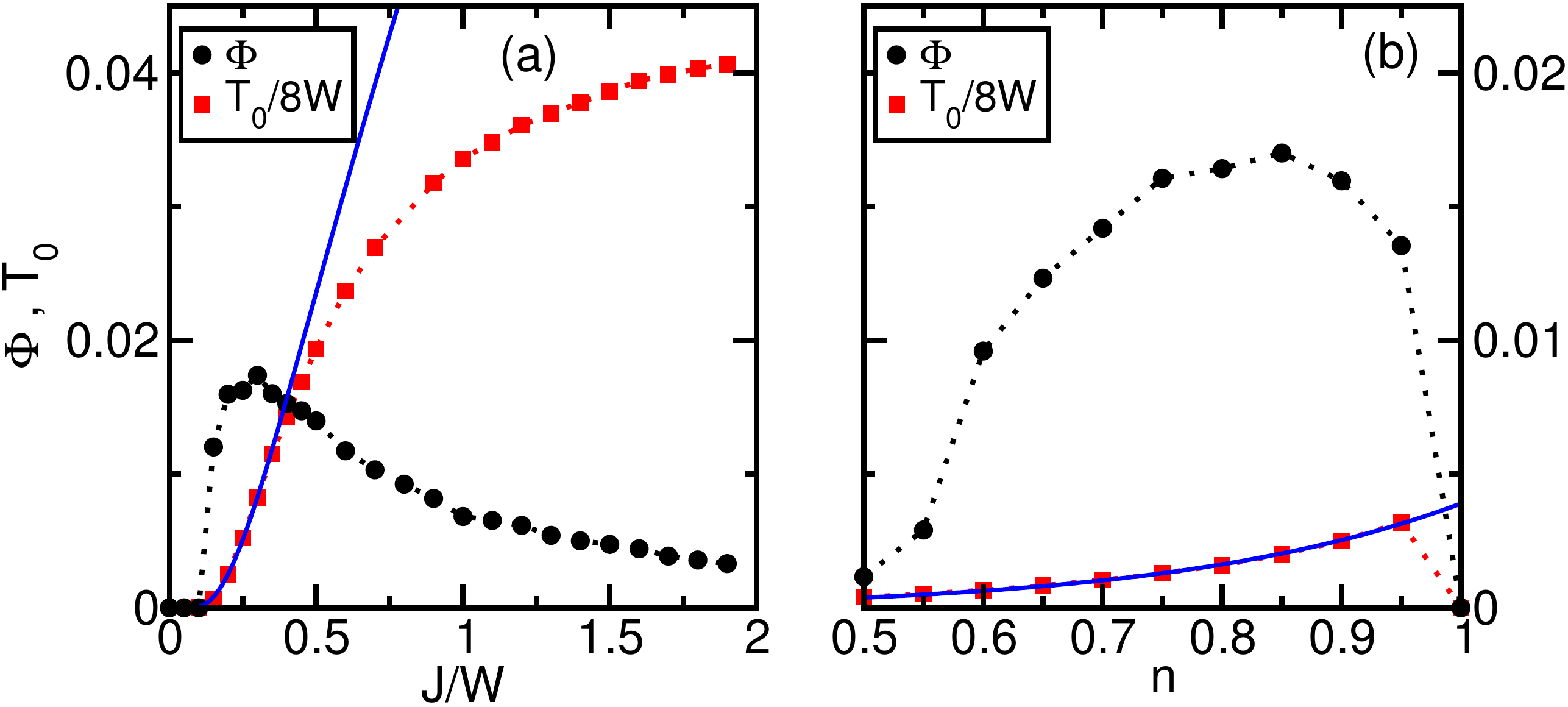}
  \caption{(color online) $\Phi$ (circles) and $T_0$ (squares) as a
  function of $J$ at fixed $n=0.9$ (a), or as a function of $n$ at fixed $J/W=0.2$ (b). The full lines
  represent approximate dependencies of $T_0$ on $J$ respectively $n$ (see text).}
\label{fig:klorder}
\end{figure}

In Fig.~\ref{fig:klorder} we  display the evolution of 
$\Phi$
along two cuts along the phase diagram indicated by the white dashed lines
in Fig.\ \ref{fig:KLPD}.
The analysis for weak Kondo coupling, $J/W<0.1$, is difficult as the signatures of SC
become very weak and hard to distinguish from numerical noise. Thus we cannot decide
whether the SC solution ceases to exist for small $J$, or whether it survives down
to $J\to0$ with an (exponentially) small pairing scale. (The latter would be expected
in the weak-coupling limit of certain mean-field theories 
\cite{suppl}.)
For $J\gtrsim W/2$, on the other hand, we observe a decay consistent with $\Phi(J)\propto 1/J$. 
We will comment on this behavior further down.


\paragraph{Normal-state Fermi-liquid scale.}
In the normal state, the KLM realizes a heavy FL at low temperatures for $n\neq
1$, with a FL (coherence) scale $T_0$ . In a local self-energy approximation, $T_0$ can be efficiently
extracted from the quasiparticle weight
\begin{equation}
 Z^{-1} =  1 -
 \left.\frac{\mathrm{d}\mathrm{Re}\Sigma(\omega)}{\mathrm{d}\omega}\right|_{\omega=0},
\end{equation}
via $T_0 = W Z$, where $\Sigma(\omega)=\Sigma_0(\omega)+\Sigma_3(\omega)$.
The evolution of $T_0$ with $J$ and $n$ is also depicted in Fig.\ \ref{fig:klorder}(a) and
(b); we recall that for $J\to0$ the scale $T_0$ depends exponentially on both $J$ and the bare $c$ DOS,
$T_0\propto \sqrt{J/W}\exp\left(-\alpha(n)\cdot W/J\right)$ with a weakly $n$-dependent coefficient $\alpha(n)$ \cite{burdin:00,pruschke:00}. For large $J\gtrsim W/2$ the dependence of $T_0$ on $J$ 
significantly deviates form this Kondo form and rather tends to saturate as $J\to\infty$.
Finally, for fixed $J$ and
varying $n$ we recover the known dependency $T_0\propto n\cdot e^{c\cdot n}$ \cite{pruschke:00}.
These different types of behavior 
for $T_0(n,J)$ can be seen from the lines superimposed to the data in Fig.\ \ref{fig:klorder}.

Apparently, there does not exist a simple connection between $T_0$ and $\Phi$. For small $J$
$\Phi(n,J)$ seems to scale with $T_0$. 
However,  as noted before, the results for very small $\Phi$ become unreliable for numerical reasons, 
i.e.\ one cannot readily extract a simple relation between $\Phi$ and $T_0$ in this limit. 
For large $J$ at $n=0.9$, on the other hand, we do not see a direct relation between $T_0$ and $\Phi$, but find 
$\Phi\propto 1/J$ instead.

\paragraph{Competition with magnetism.}
Within the DMFT for the KLM, one also finds magnetic phases, namely antiferromagnetism (AF) close to half filling and 
ferromagnetism (FM) at small filling \cite{Peters:2007,Otsuki:2009,Bodensiek:2011}. Note, however, that these
 phases have limited extent in both $J$ and $n$, for example at $n=0.9$ we find
AF only for $J<J_c(n=0.9)\approx0.2W$ \cite{Otsuki:2009}
and FM only for $n<n_c(J/W=0.2)\approx0.65$  \cite{Bodensiek:2011}. Thus, the region in 
parameter space where
we find a strong superconducting phase in Fig.\ \ref{fig:KLPD}  seems to be complementary to the regions with magnetic phases.
As the boundaries seem to overlap -- in particular the SC phase at $n=1$ lies inside the AF regime -- it is surely interesting to study the competition respectively interplay
of AF, FM and SC in detail. This is work in progress.

\begin{figure}[!b]
\includegraphics[width=0.45\textwidth]{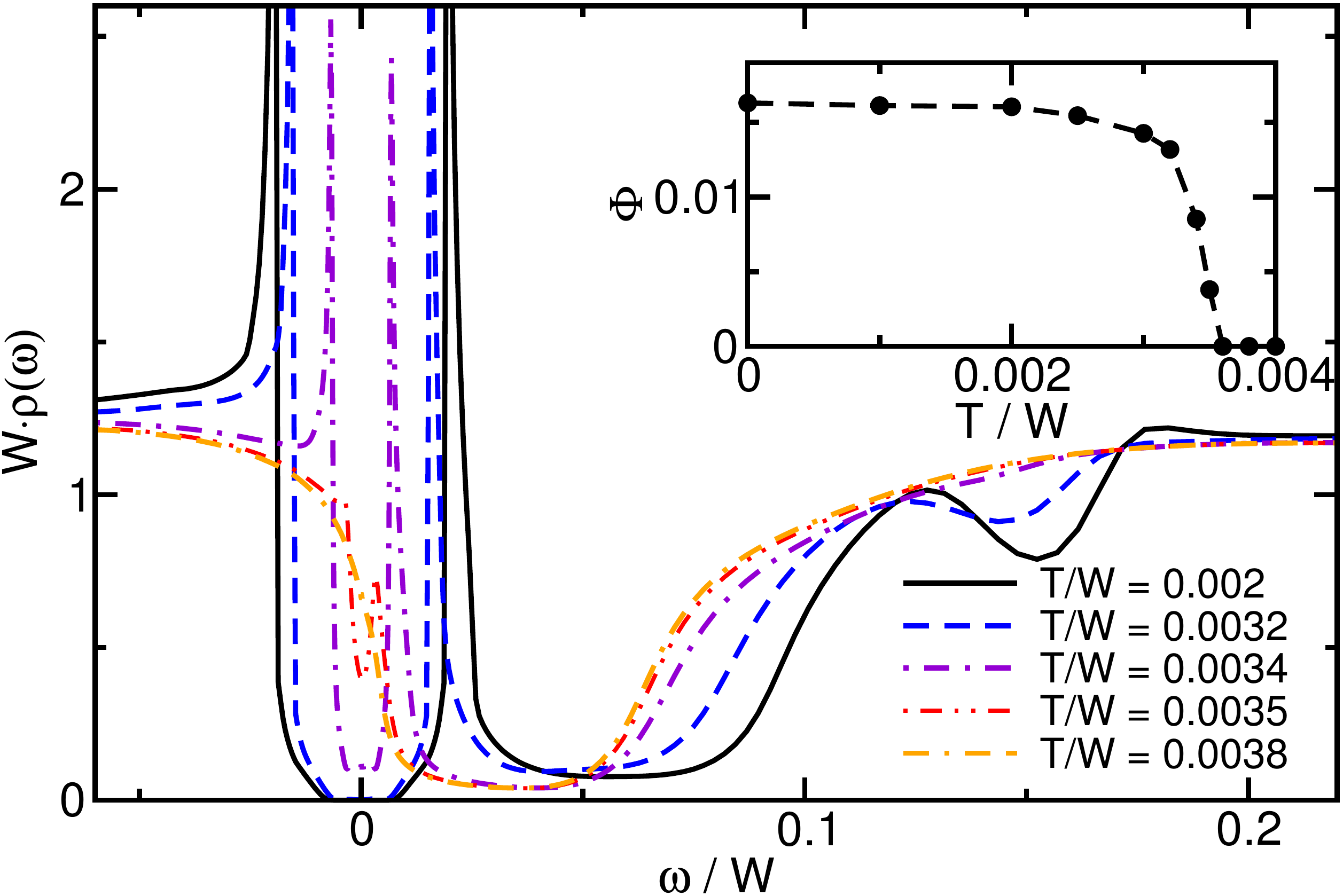}
\caption{(color online)
Evolution of the DOS with increasing temperature for $n=0.9$ and $J=0.25 W$ where $T_0/W=0.0105$ at $T\rightarrow 0$.
Inset: $\Phi$ as a function of temperature. Pair correlations are fully suppressed for $T > T_c$.
}
\label{fig:rhoT}
\end{figure}

\paragraph{Results for $T>0$.}
{Figure~\ref{fig:rhoT} displays finite-temperature results for the DOS:
With increasing temperature $T$ the gap shrinks and the spectral side resonances are
depleted.}
The reduced pair correlations are also reflected in a decrease of
$\Phi$ (see inset to Fig.~\ref{fig:rhoT}). Close to $T_c$ the gap is progressively filled, and both the
hybridization gap and the side resonances move towards the Fermi level.
Finally, in the normal-state solution for $T>T_c$ only the
hybridization pseudo-gap is visible.

\begin{figure}[!b]
\includegraphics[width=0.45\textwidth]{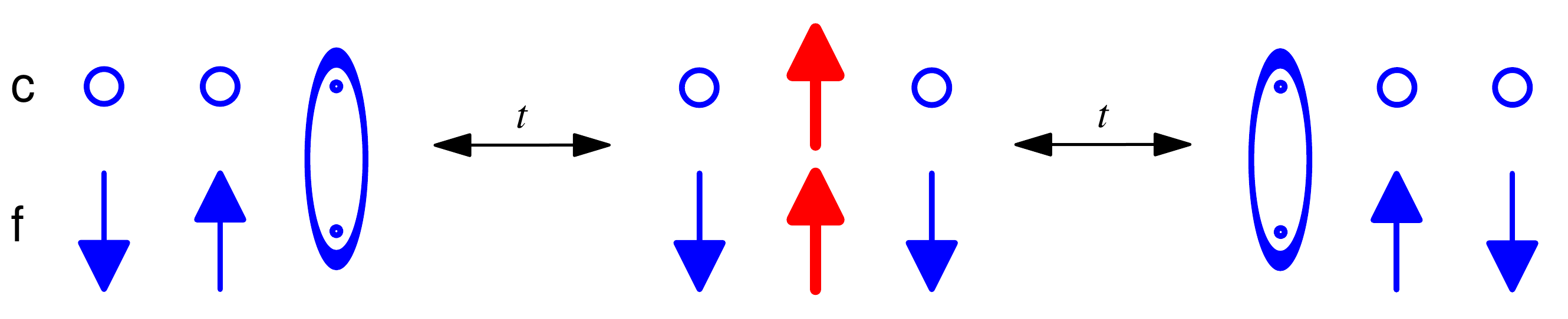}
\caption{(color online)
Second-order process responsible for pairing in the limit of large $J/W$.
Shown are the $c$ electron and local-moment (f) configurations on adjacent lattice sites;
the ellipses denote a singlet bond of two electrons.
Here, two hoppings effectively move a down-spin hole by two sites across an up-spin hole,
with a triplet intermediate state (bold/red).
}
\label{fig:config}
\end{figure}

Estimating $T_c$ from the numerical data at finite $T$ is hard, as close to the critical temperature the signatures of SC become
very small and also the convergence of the DMFT rather slow. 
From the data in Fig.\ \ref{fig:rhoT} we extract $T_c \approx 0.0036W$
at $n=0.9$ and $J=0.25 W$, which is well below $T_0$.
In Tab.\ \ref{tab:t0tc} we collect the resulting estimates for $T_c$ for fixed $n=0.9$ and several values of
$J$ in the region where according to  Fig.~\ref{fig:klorder} we have optimal conditions for pairing. 
\begin{table}[h]
\centering
\begin{tabular}{l|l|l|l|l|l}\hline
$J/W$ & 0.2 & 0.25 & 0.3 & 0.4 & 0.5\\\hline
$T_0/W$ & 0.0200 & 0.0418 & 0.0658 & 0.1139 & 0.1548\\\hline
$T_c / W$ & 0.0027 & 0.0036 & 0.0054& 0.0058 & 0.0054\\\hline
$\Phi(T=0)$ & 0.0160 & 0.0163 & 0.0174 & 0.0153 & 0.0140\\\hline
Re$\Delta(0)/W$ & 0.0138 & 0.0165 & 0.0180 & 0.0193 & 0.0195\\\hline\hline
$\text{Re}\Delta(0)/T_c$ & 5.169 & 4.583 & 3.321 & 3.305 & 3.585\\\hline
\end{tabular}
\caption{Quantities characterizing the superconducting solution for
$n=0.9$ as a function of $J$ in the region of optimal conditions for pairing.}
\label{tab:t0tc}
\end{table}
As a rule we observe that always $T_c<T_0$, i.e.\ the HF state seems to be a necessary ingredient for the
appearance of the SC phase. For small $J$ this obviously leads to a strong suppression of $T_c$. On the other hand, the gap Re$\Delta(0)$ appears to be much less sensitive
to $J$ respectively $T_0$ in this regime.   
An interesting characteristic quantity is the ratio between the gap and $T_c$. 
The results are given in the last row of Tab.\ \ref{tab:t0tc}. Obviously, the ratio exceeds the
BCS value $\Delta/T_c\approx 1.74$ by a sizeable factor between 2 and 3. 
Such values are actually observed in HF superconductors
\cite{stockert:2011}, although the interpretation there is usually given in terms of a weak-coupling
theory for a $d$-wave state. 
\paragraph{Pairing mechanism at strong coupling.}
In the strong-coupling limit, $J> W$, the pairing mechanism and behavior of $\Phi(J)$ can be understood
perturbatively, Fig.~\ref{fig:config}. We consider a conduction-band filling
of $n=N_{el}/N_s\lesssim 1$ on $N_s$ sites.
For $J/W\to\infty$ there are $N_{el}$ Kondo singlets and $(N_s-N_{el})$ uncompensated
local moments. The $c$ electrons are mobile, such that the uncompensated moments can
alternatively be interpreted as spinful $c$ holes with density $(1-n)$ and a hard-core
repulsion, forming the Fermi liquid with a coherence scale $T_0\propto W$ (
whereas the impurity Kondo scale simply diverges $\propto J$).

For large but finite $J/W$ additional excitations out of this manifold are allowed. The
lowest one consists of converting a singlet into a triplet and can be created by c
(hole) hopping. Interestingly, the excited triplet may decay via a different neighboring
hole provided that its spin is opposite to the first one. Together, this second-order
virtual process leads to correlated hopping, with an energy gain $\propto W^2/J$, which
binds two holes into a singlet state, Fig.~\ref{fig:config}. The pairing is local -- it
occurs {\it on} the site of the virtual triplet -- but the holes share this site only in
the virtual triplet state, so that the pairing is strongly retarded.

It is plausible that this pairing mechanism continues to operate at smaller $J$.
Importantly, the existence of the virtual state, whose energy may now be approximated by
the Kondo binding energy $T_0$, requires Kondo screening to be intact -- this naturally
explains the limitation $T_c<T_0$. The picture also makes clear that superconductivity is
more favorable close to half filling: Here, Kondo screening is done by $c$ electrons
whereas the paired carriers are $c$ holes (for $n<1$). In contrast, in the opposite
(exhaustion) limit of small $n$, Kondo screening becomes strongly non-local, and it is
the same $c$ electrons doing the screening that need to be paired, thus
the pairing is weak.


\paragraph{Summary.}
We have identified a novel mechanism for superconductivity in heavy-fermion materials:
Local spin fluctuations due to the Kondo exchange coupling can act as a retarded paring
interaction and drive $s$-wave superconductivity in the heavy FL.  A particularly 
interesting feature is the appearance of structures in the tunneling DOS at scales related to the
spin fluctuation spectrum, i.e.\ well separated from the coherence peaks at the gap edges. 
Such structures have been observed for example in recent STS measurements
on iron pnictides \cite{Shan:2012}. It would be interesting, to extend such STS experiments to
systematically study HF superconductors.

For model parameters relevant to HF systems, $T_c$ can be as large as several Kelvins, a
typical $T_c$ scale for existing HF superconductors. However, given that pairing in our
theory is $s$-wave, it can be ruled out for those materials where the existence of gap nodes
have been established experimentally.
More generally, it is an interesting question to what extent this mechanism can cooperate or will actually compete with SC driven by e.g.\ non-local magnetic
fluctuations.
To address this point numerical studies beyond DMFT will be required. Work along these lines is in progress.


\begin{acknowledgments}
This work was supported by the DFG through PR293/13-1 (OB,TP), the BMBF through IND 10/067 (TP) as well as FOR 960 and GRK 1621 (MV).
R\v{Z} acknowledges the support of ARRS through Program P1-0044.
NRG calculations have been done using a modification of the \textit{NRG Ljubljana} package.
Computer support was provided by the Jo\v{z}ef Stefan Institute Ljubljana and the Gesellschaft f\"ur
wissenschaftliche Datenverarbeitung G\"ottingen (GWDG).
\end{acknowledgments}


\end{document}



\title{
Supplemental Material for ``Unconventional Superconductivity from Local Spin Fluctuations in the Kondo Lattice''
}
\date{\today}
\author{Oliver Bodensiek}
\affiliation{Institut f\"ur Theoretische Physik, Universit\"at G\"ottingen, 37077
G\"ottingen, Germany}
\author{Rok \v{Z}itko}
\affiliation{J.\ Stefan Institute, Jamova 39, SI-1000 Ljubljana, Slovenia}
\author{Matthias Vojta}
\affiliation{Institut f\"ur Theoretische Physik, Technische Universit\"at Dresden, 01062 Dresden, Germany
}
\author{Mark Jarrell}
\affiliation{
Louisiana State University, Baton Rouge, Louisiana 70803, USA
}
\author{Thomas Pruschke}
\affiliation{Institut f\"ur Theoretische Physik, Universit\"at G\"ottingen, 37077
G\"ottingen, Germany}

\maketitle


\section{Local pairing in mean-field theory}

Here we summarize how local pairing emerges in the Kondo lattice model within mean-field
theory \cite{Howczak:2012,masuda:12}. One starts by representing the
f-shell local moments in terms of
auxiliary fermions:
\begin{equation}\label{sf}
\mathbf{S}_{i} = \frac{1}{2} \sum_{\sigma\sigma'}
f^\dagger_{i \sigma}
\mathbf{\tau}_{\sigma\sigma'} f_{i\sigma'}
\tag{S1}
\end{equation}
where $f_{i\sigma}$ describes a spinful (but charge-neutral) fermion destruction
operator at site $i$. This representation requires the constraint $\sum_\sigma f^{\dagger}_{i\sigma}
f_{i\sigma}=1$ and displays a local U(1) gauge invariance.

The standard decoupling of the Kondo interaction in the KLM \cite{read:83} is done
via a bosonic field $b_i$ in the particle--hole channel, resulting in an interaction term
of the form
\begin{equation}
\mathcal{H}_{b} = \sum_{i\sigma} (b_i c_{i\sigma}^\dagger f_{i\sigma} + h.c.)\,.
\tag{S2}
\end{equation}
Bose condensation of the $b_i$ endows the f particle with a physical charge and leads to
the standard picture of two hybridized electronic bands in the heavy Fermi liquid.
Alternatively, the interaction can be decoupled in the particle--particle channel,
\begin{equation}
\mathcal{H}_{\overline{b}} = \sum_{i\sigma} (\overline{b}_i c_{i\sigma}^\dagger f_{i,-\sigma}^\dagger + h.c.)\,.
\tag{S3}
\end{equation}
Importantly, condensation of the $\overline{b}_i$ alone yields the same metallic HF state,
due to the local U(1) gauge invariance.
Superconductivity is generated by a {\em simultaneous} condensation of $b_i$ and
$\overline{b}_i$ which induces anomalous expectation values among the physical c
electrons, $\langle c_{i\uparrow} c_{i\downarrow} \rangle \neq 0$, and this breaks the
physical U(1) electromagnetic symmetry  \cite{spn_note}. The resulting state displays
local $s$-wave pairing with a full gap.

This mean-field description of Kondo-lattice superconductivity via two decoupling
channels of the Kondo interaction has, however, obvious deficiencies. In particular, in
the limit $J/W\to\infty$ it yields strong pairing on the scale $J$; this result is
incorrect as the Kondo interaction only causes {\em spin} singlet formation on
this scale (but no Cooper pairing). In the
opposite limit of small $J$, pairing in this mean-field theory
becomes exponentially weak, with $T_c$ scaling as the single-impurity
Kondo temperature $T_K$.

As an aside, we note that if one starts from an Anderson (instead of Kondo) lattice
Hamiltonian, the $f$ particles always carry a physical charge, such that only a
mean-field decoupling in the particle--particle channel is required (in addition to the
c--f hybridization) to induce superconductivity. Then, the resulting pairing may be
understood as local c--f pairing \cite{Howczak:2012,masuda:12}.


\section{Local spin-fluctation theory for pairing}
Let us try to give an argument why for the KLM in DMFT the spin fluctuations can be a driving glue
for superconductivity. To this end we use the approximate form of the pairing equations as given in
\cite{Plakida}, and employ Hewson's renormalized perturbation theory
\cite{Hewson:93}, i.e.\ we assume
that we have a local Fermi liquid with correspondingly renormalized parameters. Note that this also
means that the Kondo spin is now a dynamic quantity with a fluctuation
spectrum living on the order of the Kondo scale. We do not use the
usual approximation in the Eliashberg equations that all fermionic
energies can be put to the Fermi energy, because all relevant energy
scales are of the order of $T_0$, including the spin fluctuation scale.

We are working within DMFT, i.e.\ all momentum dependencies can be ignored and the $\vec{k}$ sums
for the Green's functions be performed using the non-interacting DOS.
The equation for the gap can then be written as
$$
\Sigma^{(12)}(\omega+i0^+)=\int\limits_{-\infty}^\infty
d\omega' \,K(\omega,\omega')\;\left[-\frac{1}{\pi}\Im G^{(12)}(\omega'+i0^+)\right]
$$
where the kernel is given by
$$
K(\omega,\omega')=\frac{1}{2}\int\limits_{-\infty}^\infty d\Omega\,\frac{\tilde{g}^2\chi''_S(\Omega)}{\omega-\omega'-\Omega}\left[
\tanh\frac{\omega'}{2T}+\coth\frac{\Omega}{2T}\right].
$$
For the integral the principal value has to be taken and $\chi''_S(\omega)$ is the imaginary part of the dynamical susceptibility \emph{of the Kondo spin}.
The constant $\tilde{g}^2$ collects the
(unknown) coupling matrix element and some constant numerical factors.

At this point one usually assumes that $\omega$, $\omega'\ll\mu$ and puts $\omega\approx\omega'\approx0$, which then leads to
$$
K(\omega,\omega')\approx-\frac{\pi\tilde{g}^2}{2}\tanh\frac{\omega'}{2T}\,\chi(0)\;\;.
$$
As $\chi(0)>0$, this kernel is repulsive and hence will not generate
$s$-wave even-frequency superconductivity.
As already mentioned, we here cannot safely assume $\omega$, $\omega'\ll\mu$, since for our
Fermi liquid $\mu\to\tilde{\mu}\sim T_0$. We therefore must try to analyze the equation somewhat more carefully.
To this end we nevertheless will assume that we are deep enough in the
Fermi liquid phase so that
temperature dependencies of the Fermi liquid parameters are negligible, and that -- at least for
a determination of $T_c$ -- we can approximate $\Sigma^{(12)}(\omega+i0^+)\approx-\Delta\,\Theta(J/2-\omega)\in\mathbb{R}$. The estimate for the  cut-off is taken from the evolution of the gap function in
Fig.\ 1.
Since the gap is frequency independent, we can at least put $\omega=0$ and -- ignoring the
term with $\coth$ \cite{Plakida} 
-- find a kernel
\begin{eqnarray*}
K(0,\omega')&\approx&
\frac{\tilde{g}^2}{2}\tanh\frac{\omega' }{2T}
\int\limits_{-\infty}^\infty d\Omega\,\frac{\chi''_S(\Omega)}{-\omega'-\Omega}\\
&=&
-\frac{\pi\tilde{g}^2}{2}\tanh\frac{\omega' }{2T}\,
\chi'_S(-\omega')\\
&=&
-\frac{\pi\tilde{g}^2}{2}\tanh\frac{\omega' }{2T}\,
\chi'_S(\omega')
\end{eqnarray*}

As final input we need the anomalous Green's function, which is given by
\begin{eqnarray*}
G^{(12)}_{\vec{k}}(z)&=&\\
&&
\!\!\!\!\!\!\!\!\!\!\!\!\!\!\!\!\!\frac{\Sigma^{(12)}(z)}{(
z+\epsilon_{\vec k}-\Sigma^{(22)}(z))(z-\epsilon_{\vec k}-\Sigma^{(11)}(z))-\Sigma^{(12)}(z)^2}
\end{eqnarray*}
using the independence of the self-energy on momentum in DMFT.
We replace $\Sigma^{(12)}(z)$ by our approximation and also set it to zero in the denominator, as
we are interested in finding $T_c$. Furthermore, the diagonal self-energies are to be replaced
by the expressions for the local Fermi liquid, i.e.\
$$
\Sigma^{(\alpha\alpha)}(z)=\frac{\tilde{V}^2}{z+\tilde{\mu}}
$$
where the effective chemical potential $\tilde{\mu}$ and hybridization $\tilde{V}$ are both
O$(T_0)$. Using this self-energy expression in the normal phase leads to the well-known formation
of heavy quasiparticle bands \cite{hewson:book,Hewson:93}. In the following we further assume $\tilde{\mu}=0$.
With these simplifications we get
$$
G^{(12)}_{\vec{k}}(z)=\frac{-\Delta}{2|\epsilon_{\vec k}|}
\left[
\frac{1}{z-|\epsilon_{\vec k}|-\tilde{V}^2/z}-\frac{1}{z+|\epsilon_{\vec k}|-\tilde{V}^2/z}\right]
$$
We can now perform the $\vec k$-sum. To this end we assume a featureless DOS of the
band states with value $N_F$ and use the wide-band limit to obtain
\begin{eqnarray*}
G^{(12)}(z)
&=&
i\pi\Delta N_F\frac{1}{z-\tilde{V}^2/z}
\end{eqnarray*}

Putting everything together into the gap equation we arrive at
\begin{equation}\label{eq:Tc}
1=N_F
\int\limits_0^{J/2} d\omega\,K(0,\omega)
\left[\frac{1}{\omega-\tilde{V}}+\frac{1}{\omega+\tilde{V}}\right]
\tag{S4}
\end{equation}
as equation to determine $T_c$, where the integral has to be evaluated as principal value integral.

\begin{figure}[ht]
  \includegraphics[width=0.45\textwidth]{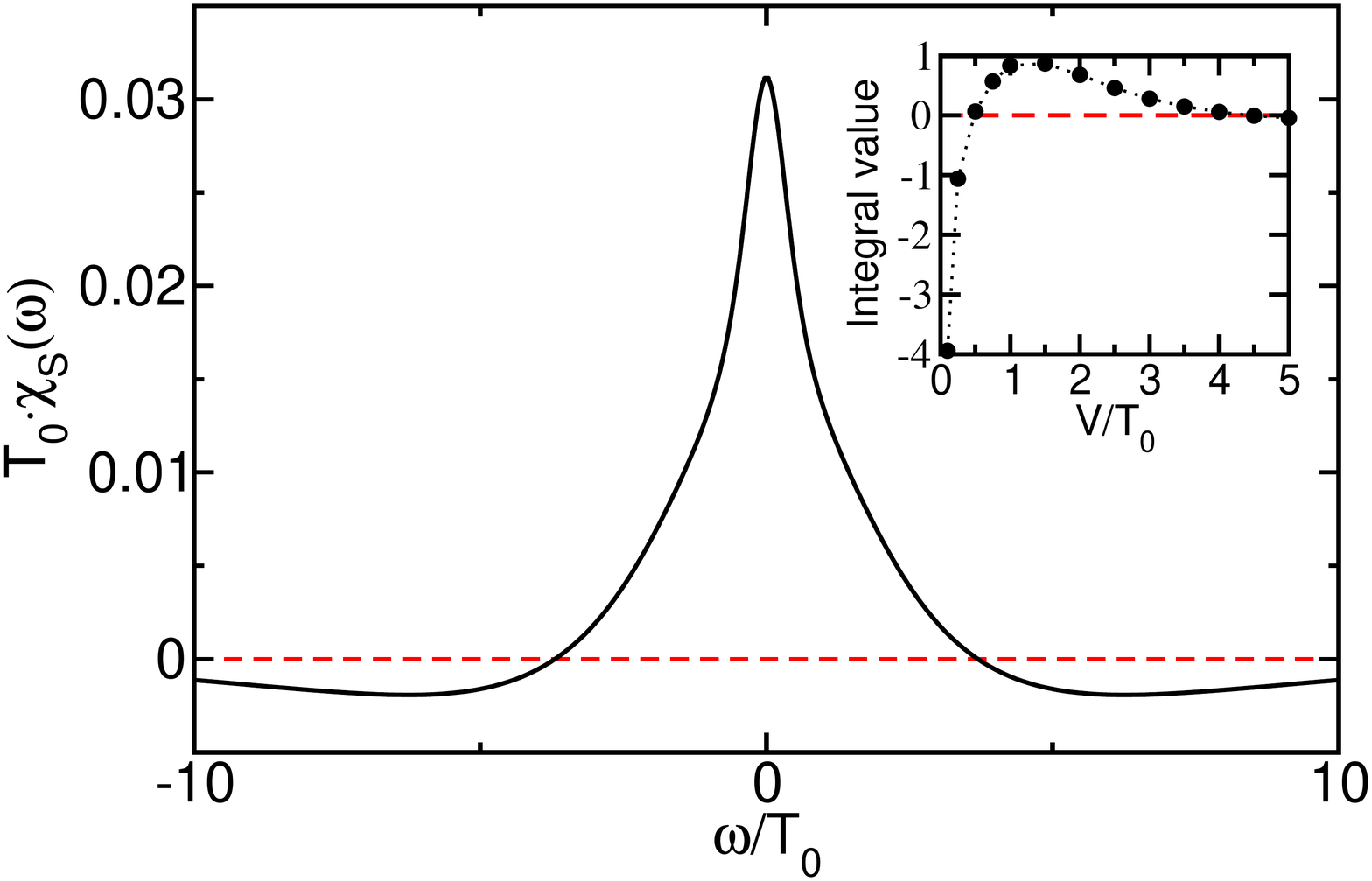}
  \caption{$\chi'_S(\omega)$ for $J=0.2W$ and $n=0.9$. The dashed line
  marks the zero. In the inset the result for the integral
  on the right hand side of Eq.~(\ref{eq:Tc}) as a function of $\tilde{V}$, evaluated for $T=0$, is shown.}
  \label{fig:chi_prime}
\end{figure}

The kernel contains $\chi_S'(\omega)$, which is shown in Fig.\ \ref{fig:chi_prime}.
Although the kernel is negative for small $\omega$ and would be considered repulsive in a
standard calculation,
the contribution from the anomalous Green's function within the approximation of a local Fermi liquid due to the Kondo effect
is also negative for energies less than some constant times $T_0$, hence the total contribution
from the region $\omega\in[0,T_0]$ is positive. There is of course the pole, which has to be
treated in the principal value sense, leading to a sign change for $\omega>\tilde{V}$. Here it becomes
important that  $\chi_S'(\omega)$ decays rapidly on a scale $T_0$, too.

As very simple approximation let us assume that $\chi_S'(\omega)\approx \chi_S(0)\,\Theta(\omega_0-|\omega|)$, where $\omega_0$ is some cut-off for the local spin fluctuations.
According to Fig.\ \ref{fig:chi_prime}, a suitable cut-off would be
again O$(T_0)$.
We restrict the discussion to $T=0$, because if we cannot find a solution there, we will
surely not find it for any $T>0$. Under these conditions the integral
on the right hand side of (\ref{eq:Tc}) then evaluates to
\begin{widetext}
$$
-\frac{\pi\tilde{g}^2\,\chi_S(0)}{2}
\int\limits_0^{\omega_0} d\omega\,
\left[\frac{1}{\omega-\tilde{V}}+\frac{1}{\omega+\tilde{V}}\right]=
-\frac{\pi\tilde{g}^2\,\chi_S(0)}{2}
\ln\left|\frac{\omega_0^2}{\tilde{V}^2}-1\right|\;\;.
$$
\end{widetext}
As long as $\omega_0<\sqrt{2}\,\tilde{V}$, the logarithm on the right hand side is negative, i.e.\ the
total expression positive and hence one has a non-trivial solution to equation  (\ref{eq:Tc}).

Let us now use the numerical data for $\chi_S'(\omega)$ and see, if the above rough arguments
are compatible with them.  As parameter we vary
the effective hybridization between some small value $\tilde{V}\ll T_0$ to some
large value $\tilde{V}\gg T_0$. The result is shown in the inset to Fig.\ \ref{fig:chi_prime}.
In both limits, the value of the integral is negative, hence no pairing is supported. However, in
a reasonably large regime with $\tilde{V}={\rm O}(T_0)$ one indeed obtains a positive value and hence the tendency to
form Cooper pairs.

One thus needs two absolutely necessary  ingredients -- hybridized bands with a small energy
scale, and a spin-fluctuation spectrum that decays on the same energy scale.
Note that this result also predicts that one has to expect a strongly reduced $T_c$ for a phonon-mediated
superconducting heavy-fermion state, as here the kernel enters with a negative sign from the outset (see e.g.\
\cite{Plakida}), and hence the low-energy part of the integral now rather tends
to suppress superconductivity.

We note that the above discussion -- considering the various uncontrolled approximations --
can neither be considered a proof of the existence of $s$-wave superconductivity, nor does
it provide a reliable formula for estimating $T_c$. For instance, we have
assumed a frequency-independent gap, obviously inconsistent with the full numerical
results in Fig.\ 1. The discussion nevertheless gives a
qualitative foundation for our results, namely that, contrary to common wisdom,
spin fluctuations can lead to $s$-wave superconductivity, at least
under the very special conditions met
in heavy-fermion systems.
